\documentclass[twocolumn,preprintnumbers,secnumarabic,amsmath,amssymb,nofootinbib,floatfix]{revtex4}
\usepackage{varioref,exscale,latexsym,amsmath,amssymb}
\usepackage{graphicx}
\usepackage{slashed}
\usepackage{graphicx}% Include figure files
\usepackage{dcolumn}% Align table columns on decimal point
\usepackage{bm}% bold math

\def\beq{\begin{equation}}

\def\eeq{\end{equation}}

\def\beqa{\begin{eqnarray}}

\def\eeqa{\end{eqnarray}}

\begin{document}

\title{{\bf Running couplings and operator mixing in the gravitational corrections to coupling constants }}

\medskip\
\author{Mohamed M. Anber${}^{1,}$ ${}^{2}$}
\email[Email: ]{manber@physics.utoronto.ca}
\author{ John F. Donoghue${}^{2}$}
\email[Email: ]{donoghue@physics.umass.edu}
\author{Mohamed El-Houssieny${}^{2,}$ ${}^{3}$}
\email[Email: ]{melhouss@physics.umass.edu}
\affiliation{${}^{1}$Department of Physics,
University of Toronto\\
Toronto, ON, M5S1A7, Canada\\
${}^{2}$Department of Physics,
University of Massachusetts\\
Amherst, MA  01003, USA\\
${}^{3}$Mathematics Department, Faculty of Science, Al-Azhar University\\
 Nasr City, Cairo, Egypt
}
\begin{abstract}
The use of a running coupling constant in renormalizable theories is well known, but the implementation of this idea for effective field theories with a dimensional coupling constant is in general less useful. Nevertheless there are multiple attempts to define running couplings including the effects of gravity, with varying conclusions. We sort through many of the issues involved, most particularly the idea of operator mixing and also the kinematics of crossing, using calculations in Yukawa and $\lambda\phi^4$ theory as illustrative examples. We remain in the perturbative regime. In some theories with a high permutation symmetry, such as $\lambda\phi^4$, a reasonable running coupling can be defined. However in most cases, such as Yukawa and gauge theories, a running coupling fails to correctly account for the energy dependence of the interaction strength. As a byproduct we also contrast on-shell and off-shell renormalizaton schemes and show that operators which are normally discarded, such as those that vanish by the equations of motion, are required for off-shell renormalization of effective field theories. Our results suggest that the inclusion of gravity in the running of couplings is not useful or universal in the description of physical processes.
\end{abstract}
%\vspace{0.2 in}
%\end{titlepage}
%\setcounter{page}{0}
%\newpage
\maketitle
%\documentstyle[12pt,epsfig]{article}
%\documentstyle[12pt,epsf,epsfig]{article}

%%%%%%%%%%%%%%%%%%%%%%%%%%%%%%%%%%%%%%%%

\section{Introduction}
Quantum corrections to scattering processes include a kinematic dependence on the energy scale of the process. In renormalizable theories, the idea of a running coupling constant absorbs a dominant and universal set of quantum corrections into a well defined logarithmic function of the energy, making this the appropriate expansion parameter in perturbation theory. Physical processes at a given energy scale are best expressed in terms of the running coupling constant defined at that scale.

Quantum corrections due to the gravitational interaction to various processes are also calculable using effective field theory methods \cite{Donoghue:1994dn}.
Because the gravitational coupling carries a dimension, the quantum corrections to a matrix element ${\cal M}$  carry a power-law dependence on the energy scale
\begin{equation}
{\cal M} \sim a \left[g + b g\kappa^2 q^2 +c g\kappa^2 q^2 \log ( -q^2)+...\right]\,,
\label{expansion}
\end{equation}
where $g$ generically denotes a coupling constant or combinations of constants and
\begin{equation}
\kappa^2 = 32\pi G_{\mbox{\scriptsize Newton}} = \frac{1}{M_P^2}\,,
\end{equation}
with $G$ being Newton's constant and $M_P = $ the Planck mass.
Following the success of running couplings in other contexts, it is tempting to try to also absorb some of the gravitational corrections into a running coupling constant $g(q^2)$. We note many such attempts  \cite{Robinson:2005fj}-\cite{Shaposhnikov:2009pv}.

It is by now well known that application of the renormalization group to effective field theories such as gravity does not lead to a traditional running coupling constant. In his influential paper on effective field theory \cite{weinberg}, Weinberg showed that the content of the renormalization group in these theories is to relate the highest powers of $q^2 \log ( -q^2/\mu^2)$ to each other. This behavior has been explored subsequently in more detail \cite{Buchler:2003vw, Bissegger:2007ib, Bijnens:2006zp, Bijnens:2009zi, polyakov}. This is due to the power counting relations of effective field theory which tell us that loop processes generate higher order operators that involve more powers of the derivatives and/or fields. Because of the increasing powers of $q^2$ in the factors of $q^{2n} \log^n ( -q^2/\mu^2)$ this does not lead to a renormalization of the leading coupling constant, but rather the  higher order couplings are renormalized. In addition, the logarithms enter differently in different processes or even in two form-factors for the same process\cite{polyakov}\footnote{Because the coefficients of the higher order operators are the ones that absorb the divergences of the effective theory, they do have a dependence on the scale $\mu$ that occurs in
dimensional regularization, but this dependence does not induce a running in the original lower order couplings.}.

Therefore attempts to define a running coupling necessarily involve definitions which fall outside of the usual renormalization group. They tend to involve attempts to identify a typical $q^2$ in an amplitude with a renormalization scale\footnote{In this paper we will use the notation $E$ or $M$ for various definitions of the renormalization scale and reserve $\mu$ for the scale that arises in dimensional regularization through the factor $\mu^{4-d}$ in front of dimensionally regularized loop integrals.} $M^2$ and absorbing the higher order $q^2$ effects into the coupling constant
\begin{equation}
g(M^2) = g + b g\kappa^2 M^2 +c g\kappa^2 M^2 \log ( M^2)\,.
\label{generic expansion}
\end{equation}
It is worth exploring whether such a definition can make sense in physical processes.

This procedure is also outside of the usual effective field theory methods. One applies effective field theory as an low energy expansion about zero energy. Higher order energy dependence is associated with operators that are higher order in the energy expansion\footnote{If particle masses are considered, there is some renormalization of low order operators, however, we consider massless theories in this paper.} and does not renormalize the low order operators of the theory. A coupling constant definition such as Eq. \ref{generic expansion} attempts to define the theory around the high energy renormalization scale M. In this case, success could be achieved if the definition correctly captures the effects of the loop corrections at this energy scale.

Some potential pitfalls are visible in this strategy.
\begin{itemize}
\item One is that in the effective field theory context, higher momentum dependence is associated with new operators carrying extra derivatives, not the original operator carrying the coupling $g$. This brings in the need for {\em operator mixing}, as renormalization conditions at a given scale can involve a mixture of a set of operators.

\item However, even if one can define a running coupling using some combination of the relevant operators, there is the question of {\em universality}. That is, one has to assess whether the operator mixing is the same for all processes. It is possible that a definition that is useful for one reaction may be deleterious when used in another reaction.

\item In addition, there is a problem of {\em kinematics} and {\em crossing}, in that the energy variable $q^2$ can take on different values and even different signs in different contexts. For example, in space-like versus time-like reactions, $q^2$ changes sign, so that a running coupling that decreases with energy for a space-like process will increase with energy if that same reaction is crossed into the time-like regime. Crossing will turn out to be a major obstacle in the Yukawa and gauge theory cases.

\end{itemize}

In this paper we will explore these issues using calculations of the gravitational corrections to $\lambda \phi^4$ and Yukawa theories. We use these theories as test cases in order to avoid the irrelevant complications of gauge invariance. However the general lessons of our results will apply to other theories also. We will find that outside of some special cases, the idea of the gravitational contribution to the running of a coupling constant is not a useful idea in the perturbative regime.

\section{Preview of key issues}

Let us first review various ways of calculating the running coupling in renormalizable theories.

At the most physical level, one can calculate any given physical process including quantum corrections and identify the large logarithms that can be absorbed into the running coupling. However, the use of physical processes can sometimes be complicated by the presence of imaginary parts to the amplitudes and by not being sure what part of the quantum corrections are universal enough to be absorbed in the coupling constant.

To address these issues, one can alternatively define the coupling by renormalizing at an unphysical Euclidean point $p^2 =-M^2$, avoiding the cuts and poles of the physical amplitudes.

A third method of great practical utility is to study the divergences of the coupling constant. In dimensional regularization, the $1/\epsilon$ divergences are always accompanied by $\log \mu$, where a factor of $\mu^\epsilon$ is introduced to keep the dimensionality of loop integrals unchanged. By dimensional analysis then, if the only large scale in the theory is the renormalization scale $M$, the $\log \mu$ dependence always tracks the $\log M$ dependence. The $1/\epsilon$ behavior is obviously universal, since it goes into the renomalization of the coupling constant, and the accompanying $\log M$ is also universal and can be readily incorporated into the running coupling constant $g(M)$. This is powerfully exploited in renormalization group arguments to show that this running coupling is the appropriate coupling for all processes at this scale.

If we look at the nature of the gravitational corrections, we see some crucial differences. The well-known presence of divergences in gravitational loops is not itself a significant issue, but the nature of the renormalization procedure dealing with them is important. Because the gravitational coupling carries inverse mass dimensions, the divergences go into the renormalization of new operators that carry extra derivatives. For example, when we discuss the Yukawa couplings of a scalar $\phi$ and fermion $\psi$, originally of the form
\begin{equation}
{\cal L}_Y =  - g \phi \bar{\psi}\psi\,.
\end{equation}
At low energy, the divergences go into the renormalization of the coefficients of the higher order operator
\begin{equation}
{\cal O}_{g_3}=g_3\phi\partial_{\mu}\bar\psi\partial^{\mu}\psi\,.
\end{equation}
or, as we discuss below, into higher dimension four-fermion operators.
Following the $1/\epsilon$ behavior of the loops will {\em not} capture the renormalization of the original operator.

Even if we give up this useful technique, we can still study the energy dependence of physical processes or consider the strength of the interaction at a Euclidean point $p^2=-M^2$. It is clear that this will then involve a linear combination of the initial operator plus the higher order operators. This is what we mean by operator mixing. A rule for such a procedure can always be developed to define a coupling at a given energy and this will yield a given definition of a running coupling that includes power law running. But the question then is whether this definition is useful. To be useful, it should be in some sense universal, so that it applies to other reactions also, and it should encapsulate at least some of the large corrections to physics processes. The presence of multiple operators in the effective field theory basis argues against universality - different operators contribute differently from process to process. And power-law kinematics also argues against utility, because as mentioned previously a kinematic variable that is positive in one reaction is negative in a related reaction. A coupling that minimizes the energy dependence of one process with increase the energy dependence of the related process.\footnote{Note that the sign of the momentum is not an issue with a running coupling with logarithmic behavior, $\log (-|q^2|) = \log (|q^2|) +i\pi$. The magnitude of the logarithm is universally present in both spacelike and timelike processes, while the imaginary part is part of the residual quantum correction.}

It is possible that this procedure can still work. In $\lambda \phi^4$ theory we will see that physical processes regularly involve a mixture of spacelike and timelike subdiagrams, because of the permutation symmetry of the original interaction. In this case, we will be able to define a reasonable running coupling with power law running, subject to only modest ambiguities due to the renormalization scheme.

However, in most theories we find that the power law running is not a useful concept. In the perturbative regime, these theories are better described by an operator basis with coupling constants that do not run nor mix.

%%%%%%%%%%%%%%%%%%%%%%%%%%%%%%%%%%%%%%%%%%%%%%%%%%%%%%%%%%%%%
\section{Gravitational Corrections to $\lambda\phi^4$ Interaction}
%%%%%%%%%%%%%%%%%%%%%%%%%%%%%%%%%%%%%%%%%%%%%%%%%%%%%%%%%%%%%

In this section we will explore the various ways of defining a running coupling in $\lambda\phi^4$ theory. This effort is reasonably successful, and provide an illustration of what powerlaw running could look like. The feature that is most important in this construction is the mix of spacelike and timelike diagrams, with a high permutation symmetry.

We consider a massless real scalar $\phi$ with a $\lambda\phi^4$ interaction, coupled minimally to gravity. The Lagrangian reads
\begin{eqnarray}
\nonumber
\sqrt{-g}{\cal L}&=&\frac{2}{\kappa^2}\sqrt{-g}R \nonumber \\&&+\sqrt{-g}\left[\frac{1}{2}g^{\mu\nu}\partial_{\mu}\phi\partial_{\nu}\phi
-\frac{\lambda}{4!}\phi^4\right]\,,
\end{eqnarray}
where $\kappa^2=32\pi G_{\mbox{\scriptsize Newton}}$, $g^{\mu\nu}$ is the metric tensor, $R$ is Ricci scalar, and $\lambda$  is the scalar-self coupling.

Temporarily ignoring the gravitational interaction, the one loop scattering amplitude in this theory is derived from the diagrams of Fig. \ref{Scalar} (a), (b), and has the form
\begin{eqnarray}
\nonumber
-i{\cal M}&=& -i\lambda + \frac{3i\lambda^2}{32\pi^2}\left[ \frac{2}{\epsilon} +\log 4\pi -\gamma\right] \\
\nonumber &&-\frac{i\lambda^2}{2\left(4\pi\right)^2}\left[\log\left(\frac{-s}{\mu^2}\right)+\log\left(\frac{-t}{\mu^2}\right)\right.\\
\label{scattaringscalarnogravity}
&&\left.+\log\left(\frac{-u}{\mu^2}\right)\right]\,,
\end{eqnarray}
where $\epsilon=4-d$ and $\lambda$ is the bare coupling constant, and the channels $s$, $t$ and $u$ are defined as usual $s=(p_1+p_2)^2$, $t=(p_1-p_3)^2$, $u=(p_2-p_3)^2$. In order to use the on-shell process to define a running coupling, we can choose to measure the renormalized coupling at the point $s=2E^2$, $t=u=-E^2$. This lets us define the effective coupling constant $\lambda(E)$ as
\begin{eqnarray}
\nonumber
-i\lambda(E)&=&-i\lambda+ \frac{3i\lambda^2}{32\pi^2}\left[ \frac{2}{\epsilon} + \log 4\pi -\gamma\right]\\
\label{scalar on-shell scattering}
&&-\frac{i\lambda^2}{2\left(4\pi\right)^2}\left[\log\left(\frac{2E^2}{\mu^2}\right)+2\log\left(\frac{E^2}{\mu^2}\right)\right]
\end{eqnarray}
such that the on-shell perturbative scattering amplitude becomes
\begin{eqnarray}
\nonumber
-i{\cal M}= -i\lambda(E)&-&\frac{i\lambda^2(E)}{2\left(4\pi\right)^2}\left[\log\left(\frac{s}{2E^2}\right)+\log\left(\frac{-t}{E^2}\right)\right.\\
\label{scattaringscalarnogravity}
&&\left.+\log\left(\frac{-u}{E^2}\right) +i\pi\right]\,.
\end{eqnarray}
For all $s,~t,~u$ of order $E^2$ all the logarithms are small. The potentially large logs have been absorbed into $\lambda(E)$. The quantum corrections proportional to $\lambda^2$ vanish at the renormalization point and are small throughout the physical region. The $\beta$ function is calculated from
\begin{equation}\label{beta function on shell no gravity}
\beta(\lambda)\equiv E\frac{\partial \lambda(E)}{\partial E}=\frac{3\lambda^2}{16\pi^2}\,.
\end{equation}

It is often common to renormalize a symmetric off-shell Euclidean point. In this case we treat all lines as incoming and choose kinematics $p_i^2= -M^2, ~~ s=t=u=-4M^2/3$. This allows a definition
\begin{eqnarray}
\nonumber
-i\lambda(M)&=&-i\lambda + \frac{3i\lambda^2}{32\pi^2}\left[ \frac{2}{\epsilon} +\log 4\pi -\gamma\right]\\
\label{scalar off-shell scattering }
&&- \frac{3i\lambda^2}{2\left(4\pi\right)^2}\log\left(\frac{4M^2}{3\mu^2}\right),
\end{eqnarray}
with a similar expansion of the matrix element, and related beta function. Finally, we note how the beta function can be read off from the coefficient of the $1/\epsilon$, since it is intrinsically connected to the $\log \mu$ in the scattering amplitude.

We now turn to the comparable definitions of the running coupling with the inclusion of gravitational corrections. We employ two different methods. In all methods, we use dimensional regularization scheme to regularize our integrals. First, we define the running coupling constant as an effective coupling for scattering processes. In the second method we  use an {\em off-shell} procedure to calculate the $\beta$ function. These will yield similar results.

%%%%%%%%%%%%%%%%%%%%%%%%%%%%%%%%%%%
\subsection{Gravitational corrections in on-shell scattering processes}
%%%%%%%%%%%%%%%%%%%%%%%%%%%%%%%%%%%

Let us now include gravity. To find the graviton propagator one perturbs the metric tensor about the flat background $g_{\mu\nu}=\eta_{\mu\nu}+\kappa h_{\mu\nu}$, where $h_{\mu\nu}$ are the spacetime fluctuations. Then, one expands $R$ in terms of $h_{\mu\nu}$ and writes $R$ in the form $h_{\mu\nu}O^{\mu\nu,\rho\sigma}h_{\rho\sigma}$. To simplify the calculations, we fix the gauge freedom by employing the harmonic gauge $\partial^{\mu}h_{\mu\nu}-\partial_{\nu}h^{\alpha}_{\alpha}/2=0$. Finally, one obtains the graviton propagator
\begin{equation}
D_{\mu\nu,\rho\sigma}(q^2)=i\frac{-\frac{1}{2}\eta_{\mu\nu}\eta_{\rho\sigma}+\frac{1}{2}\eta_{\mu\sigma}\eta_{\rho\nu}+\frac{1}{2}\eta_{\mu\rho}\eta_{\nu\sigma}}{q^2}\,.
\end{equation}

\begin{figure}[ht]
\leftline{
\includegraphics[width=.45\textwidth]{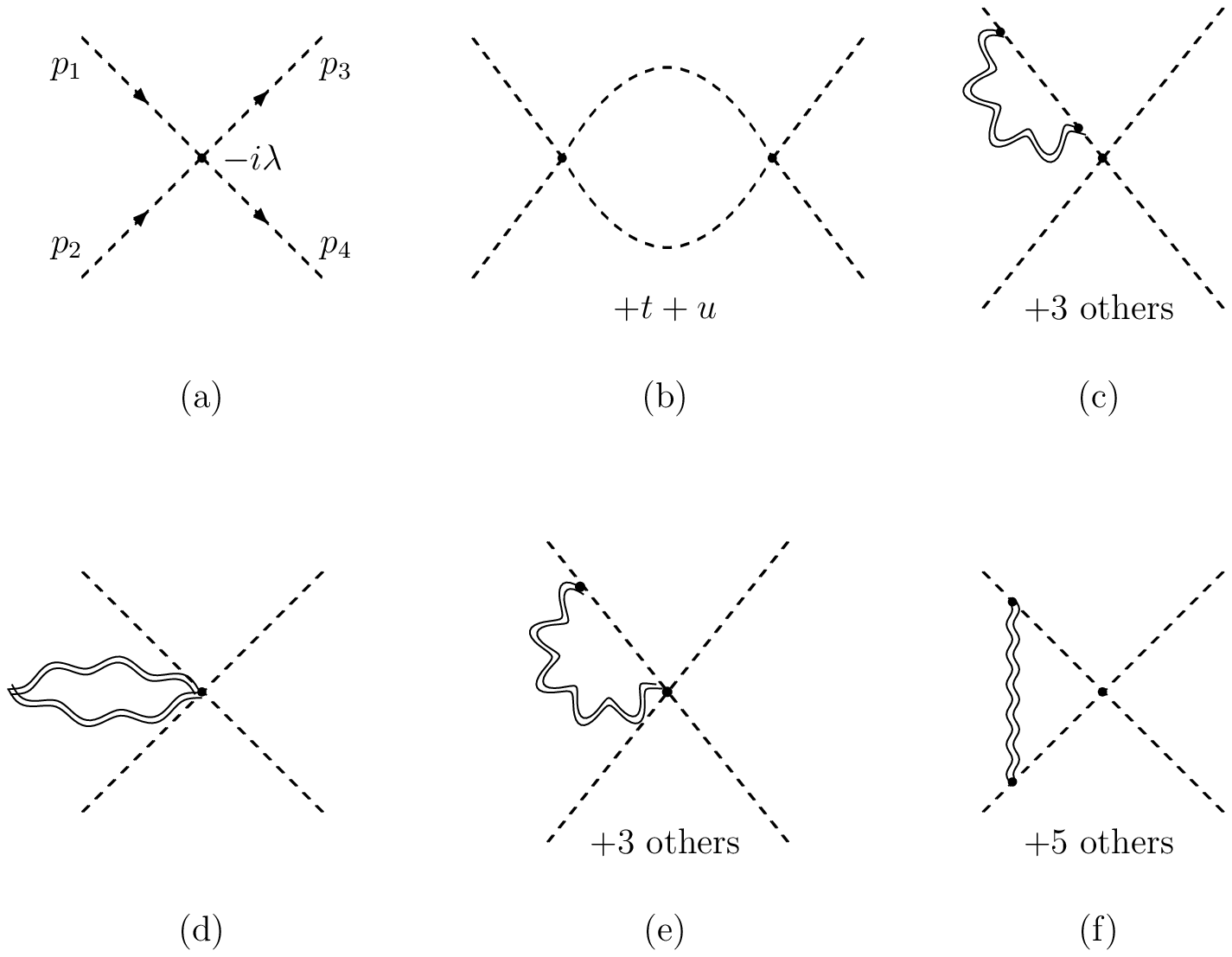}
}
\caption{The contributing diagrams to the running of the scalar coupling constant $\lambda$.}
\label{Scalar}
\end{figure}

We will first explore an {\em on-shell} renormalization scheme. We consider the different diagrams contributing to the on-shell scattering process $\phi+\phi\rightarrow \phi+\phi$. In addition to the quantum scalar corrections to the $s$, $t$ and $u$ channels (Fig. \ref{Scalar} (b)), we include the gravitational corrections to the wave-function and vertex, as shown in Fig. \ref{Scalar} (c), (d), (e) and (f). The gravitational wave-function renormalization and diagram (e) vanish for massless particles. Also, diagram (d) does not contribute to the vertex corrections in the dimensional regularization scheme. Hence, we are left only with diagram (f)
\begin{eqnarray}
\nonumber
{\cal A}_{(f)}&=&-\frac{i\kappa^2\lambda}{2\left(4\pi\right)^2}\left[s\log\left(\frac{-s}{\mu^2}\right)+t\log\left(\frac{-t}{\mu^2}\right)\right.\\
\nonumber
&&\left.+u\log\left(\frac{-u}{\mu^2}\right)\right]\\
\label{scattaringscalar}
&&+\frac{i\kappa^2\lambda}{2\left(4\pi\right)^2}\left[s^2C(s)+t^2C(t)+u^2C(u) \right]\,,
\end{eqnarray}
where $\mu$ is introduced in the scalar vertex as $\mu^{4-d}\lambda \phi^4/4!$ for the purpose of dimensional regularization.
The function $C(x)$ is given by the integral
\begin{equation}\label{IR integral}
C(x)=\int_0^1\int_0^1 dw d\xi\frac{1}{w\xi(1-\xi)x}\,.
\end{equation}
These integrals are IR divergent which makes their inclusion problematic in calculating the scattering amplitudes. In principle, one needs to include as well the scalar masses and the contribution from soft gravitons that potentially will remove the IR divergence. We do not follow this procedure here. Instead, we choose to ignore altogether the contributions from $C(x)$. Another procedure that avoids these divergences will be followed in the next sections, where we perform all the calculations {\em off-shell}.

%%%%%%%%%%%%%%%%%%%%%%%%%
%%%%%%%%%%%%%%%%%%%%%%%%%%%%%%

Note that there is no UV divergence in the above on-shell scattering amplitude. This is a special feature of $\phi^4$ theory and arises from the permutation symmetry of the Feynman diagrams. While individual diagrams are of course divergent, the divergent pieces sum to
\begin{equation}
{\cal M}_{div}\sim  \lambda\kappa^2\frac{1}{\epsilon}(s+t+u) \, .
\label{divergences stu}
\end{equation}
However this sums to zero since we have
\begin{equation}
s+t+u=0
\label{stu}
\end{equation}
when evaluated on-shell\footnote{If we had used massive scalars, we would obtain UV poles $\sim m^2(1/\epsilon-\log(p^2/\mu^2)+\mbox{finite})$.}. Note also the related feature that despite the apparent presence of $\mu$ in the scattering amplitude Eq. \ref{scattaringscalar}, in fact this amplitude is independent of $\mu$ because of the on-shell identity Eq. \ref{stu}.

A related unique feature for this theory is that the higher order operator vanishes by the equations of motion, or equivalently can be removed by a field redefinition. The higher order operator that is generated by one loop gravitational corrections is
\begin{equation}
{\cal L}_{\lambda_1}\equiv -\lambda_1\phi^2\partial_{\mu}\phi\partial^{\mu}\phi \, .
\label{higheroperator}
\end{equation}
This would generate a matrix element proportional to $s+t+u$, which as we have seen vanishes. By integrating it by parts it can be seen to be equivalent to the operator $\phi^3 \Box \phi$, which vanishes by the equation of motion. Such operators can be removed from the operator basis in favor of other local operators. In this case we can see this by performing the field redefinition
\begin{equation}
 \phi' =\phi - \frac{\lambda_1}{3} \phi^3\,,
\end{equation}
which removes the term ${\cal L}_{\lambda_1}$ and to first order in $\lambda_1$ generates the operator $\sim\lambda\lambda_1 \phi^6$, which comes from the expansion of the original $\phi^4$ interaction. In dimensional regularization this $\phi^6$ operator does not mix with $\phi^4$ at one loop since the massless tadpole loop integral vanishes. Therefore for onshell renormalization we do not need to consider operator mixing at this order. We will however return to this issue in the next section.

Now, we choose to renormalize (\ref{scattaringscalar}) at the physical point $s=2E^2$, $t=u=-E^2$, and then define the effective coupling constant $\lambda(E)$ as
\begin{equation}\label{scalar on-shell scattering}
-i\lambda(E)=-i\lambda-\frac{3i\lambda^2}{2\left(4\pi\right)^2}\log\left(\frac{E^2}{\mu^2}\right)-\frac{i\log(2)\kappa^2\lambda}{\left(4\pi\right)^2}E^2,
\end{equation}
where the second term is the quantum scalar correction, and we have omitted an imaginary phase. Now the full amplitude is given by
\begin{eqnarray}
\nonumber
{\cal A}_{(f)}&=&-i\lambda(E)-\frac{i\lambda^2(E)}{2\left(4\pi\right)^2}\left[\log\left(\frac{s}{2E^2}\right)+\log\left(\frac{-t}{E^2}\right)\right.\\
\nonumber
&&\left.+\log\left(\frac{-u}{E^2}\right) +i\pi\right] \\
\nonumber
&&
-\frac{i\kappa^2\lambda (E)}{2\left(4\pi\right)^2}\left[s\log\left(\frac{-s}{2E^2}\right)+t\log\left(\frac{-t}{E^2}\right)\right.\\
\nonumber
&&\left.+u\log\left(\frac{-u}{E^2}\right)\right]\\
\label{scattaringtotal}
&&+\frac{i\kappa^2\lambda(E)}{2\left(4\pi\right)^2}\left[s^2C(s)+t^2C(t)+u^2C(u) \right]\,,
\end{eqnarray}
This has been a useful definition as the hard quantum corrections for the gravitation loops vanish at the renormalization point and stay small in the physical region.

Using this definition, the $\beta$ function reads
\begin{equation}\label{beta function on shell}
\beta(\lambda)\equiv E\frac{\partial \lambda(E)}{\partial E}=\frac{3\lambda^2}{16\pi^2}+\frac{\log(2)\kappa^2\lambda}{8\pi^2}E^2\,.
\end{equation}
Under this procedure, the gravitational corrections do not tend towards an asymptotically free theory while in the perturbative region.

%%%%%%%%%%%%%%%%%%%%%%%%%%%%%%%%%%%%%%%%%%%%%%%%%%%%%%%%%%%%%%%%%
\subsection{Off-shell renormalization}
%%%%%%%%%%%%%%%%%%%%%%%%%%%%%%%%%%%%%%%%%%%%%%%%%%%%%%%%%%%%%%%%%

In this section, we provide another method to calculate the $\beta$ function by using an off-shell renormalization point. In renormalizable theories, going off-shell provides no essential complication, because we know all the operators that can be renormalized. It actually provides a simplification, as we can choose a convenient symmetric point, and in the Euclidean region can avoid poles and cuts. However, in our case, the off-shell point brings in a potential complication, as new divergences appear and the higher order operator of Eq. \ref{higheroperator} now gives a non-vanishing matrix element. These two issues are related, as the divergence is absorbed in the coefficient of the higher order operator. Thus we have operator mixing appearing at the off-shell point, while it did not appear on-shell. However, by dealing with this feature we can still obtain results similar to the on-shell running coupling.

In this method we compute the scattering amplitude using off-shell momenta. In this case, the diagrams of Fig. \ref{Scalar} generate the operator ${\cal O}_{\lambda_1}=-\lambda_1\phi^2\partial^\mu\phi\partial_\mu\phi/8$. The scattering amplitude is given by
\begin{eqnarray}
\nonumber
{\cal A}&=&-i\lambda-\frac{i\lambda_1}{4}\sum_ip_i^2+\frac{3i\lambda^2}{2\left(4\pi\right)^2}\left[\frac{2}{\epsilon}+\log4\pi-\gamma \right]\\
\nonumber
&&+\frac{i\kappa^2\lambda(s+t+u)}{2\left(4\pi\right)^2}\left[\frac{2}{\epsilon}+\log4\pi-\gamma \right]\\
\nonumber
&&-\frac{i\lambda^2}{2\left(4\pi\right)^2}\left[\log\left(\frac{-s}{\mu^2}\right)+\log\left(\frac{-t}{\mu^2}\right)+\log\left(\frac{-u}{\mu^2}\right) \right]\\
\nonumber
&&-\frac{i\kappa^2\lambda}{2\left(4\pi\right)^2}\left[s\log\left(\frac{-s}{\mu^2}\right)\right.\\
\nonumber
&&\left.+t\log\left(\frac{-t}{\mu^2}\right)+u\log\left(\frac{-u}{\mu^2}\right)\right]+{\cal Z}\,,
\end{eqnarray}
where
\begin{eqnarray}
\nonumber
{\cal Z}=
&&\frac{i\kappa^2\lambda}{4\left(4\pi\right)^2}\left[\left(s^2-(p_1^2+p_2^2)s \right)C(p_1^2,p_2^2,s) \right.\\
\nonumber
&&\left.+\left(t^2-(p_1^2+p_3^2)t \right)C(p_1^2,p_3^2,t)  \right.\\
\nonumber
&&\left. +\left(u^2-(p_2^2+p_3^2)u \right)C(p_2^2,p_3^2,u)\right.\\
\nonumber
&&\left.+\left(s^2-(p_3^2+p_4^2)s \right)C(p_3^2,p_4^2,s)\right.\\
\nonumber
&&\left.+\left(u^2-(p_1^2+p_4^2)u \right)C(p_1^2,p_4^2,u)\right.\\
&&\left.+\left(t^2-(p_2^2+p_4^2)t \right)C(p_2^2,p_4^2,t)\right]\,,
\end{eqnarray}
and, for example,
\begin{eqnarray}
\nonumber
&&C(p_1^2,p_2^2,s)=\\
\nonumber
&&\int_{0}^1\int_{0}^1
\frac{dwd\xi}{(1-w)\left( p_1^2(1-\xi)+p_2^2\xi\right)+w\xi(1-\xi)s}\,.\\
\end{eqnarray}
We then compute the scattering amplitude at the Euclidean momenta $p_1^2=p_2^2=p_3^2=p_4^2=-M^2$, and $s=t=u=-4M^2/3$. Hence, we find
\begin{eqnarray}
\nonumber
{\cal A}&=&-i\lambda+iM^2\lambda_1\\
\nonumber
&&+\frac{3i\lambda^2}{2\left(4\pi \right)^2}\left[\frac{2}{\epsilon}+\log 4\pi-\gamma-\log\left(\frac{4M^2}{3\mu^2}\right)\right]\\
\nonumber
&&+\frac{i\kappa^2\lambda M^2}{\left(4\pi\right)^2}\left[-\frac{4}{\epsilon}-2\log 4\pi+2\gamma\right.\\
&&\left.+2\log\left(\frac{4M^2}{3\mu^2}\right)+\rho \right]\,,
\end{eqnarray}
where $\rho=-i(4\pi)^2{\cal Z}/\kappa^2\lambda M^2$ is some numerical coefficient whose value does not affect the $\beta$ function
\begin{equation}
\rho=4\int_{0}^1d\xi\frac{\log \left[4\xi(1-\xi)/3\right]}{-3+4\xi-4\xi^2}\approx 2.83\,.
\end{equation}

In order to deal with the issue of operator mixing, we need to choose appropriate renormalization conditions. Now, we define effective couplings $\lambda(M)$, and $\lambda_1(M)$ such that the scattering amplitude is given by
\begin{equation}\label{first condition}
{\cal A}=-i\lambda(M)+iM^2\lambda_1(M)\,.
\end{equation}
For the higher order operator we define
\begin{equation}\label{second condition}
\frac{\partial {\cal A}}{\partial M^2}=i\lambda_1(M)\,.
\end{equation}
Because $\lambda_1$ is generated through loops and is higher order in  the energy expansion, we treat it as a quantity of order $\lambda^2$ or $\lambda\kappa^2$. This allows us then to neglect any feedback from $\lambda_1$ into the original coupling $\lambda$. Such feedback would occur from further loops and would be of higher order $\lambda_1\lambda\sim{\cal O}(\lambda^3)$ or $\lambda_1\kappa^2\sim{\cal O}(\lambda^2\kappa^2)$, both of which we drop.

Using these definitions, the divergence $\sim 1/\epsilon$ is absorbed as usual in $\lambda$, and  the new divergence $\sim (s+t+u)/\epsilon$ is absorbed into $\lambda_1$. We then solve the system (\ref{first condition}) and (\ref{second condition}) simultaneously to find
\begin{eqnarray}
\nonumber
\lambda(M^2)&=&\lambda+\frac{3\lambda^2}{2\left(4\pi\right)^2}\left[-1-\log4\pi+\gamma\right.\\
\nonumber
&&\left.+\log\left(\frac{4M^2}{3\mu^2}\right) \right]
+\frac{2\kappa^2\lambda M^2}{\left(4\pi\right)^2}\,,\\
\nonumber
\lambda_1(M^2)&=&\lambda_1-\frac{3\lambda^2}{2\left(4\pi\right)^2M^2}\\
\nonumber
&&+\frac{2\kappa^2\lambda}{\left(4\pi\right)^2}\left[1-\log 4\pi+\gamma+\frac{\rho}{2}+\log (4/3)\right.\\
&&\left.+\log\left(\frac{M^2}{\mu^2}\right) \right]\,.
\end{eqnarray}

Finally, the $\beta$ functions of $\lambda$ and $\lambda_1$ read
\begin{eqnarray}
\nonumber
\beta(\lambda)&\equiv& M\frac{\partial \lambda(M)}{\partial M}=\frac{3\lambda^2}{\left(4\pi\right)^2}+\frac{\kappa^2\lambda}{4\pi^2}M^2\,,\\
\beta(\lambda_1)&\equiv& M\frac{\partial \lambda_1(M)}{\partial M}=\frac{3\lambda^2}{\left(4\pi\right)^2M^2}+\frac{\kappa^2\lambda}{4\pi^2}\,.
\end{eqnarray}

\subsection{Lessons from $\phi^4$ theory}

We have used two different methods to define the running coupling. Both on-shell and off-shell Euclidean methods yielded similar results. The higher order operator in this situation vanishes for on-shell matrix elements, so that the on-shell method did not require any operator mixing and resulted in a finite amplitude. However we needed this unphysical operator and operator mixing in order to accomplish the renormalization when working off-shell. Of course when we continue and apply this operator to a physical process it will again vanish. The slight differences in the beta functions can be accommodated by a scheme dependent renormalization scale.

In renormalizable theories, we also know that the running coupling will also be the one relevant for loop diagrams at a given energy. One way to show this is to use renormalization group arguments. However, we also know this from the structure of Feynman diagrams. While loop momenta run over all energies, in dimensional regularization the loop integrals are dominated by the overall energy scale of the problem, aside from infrared and collinear regions that can be dealt with using other means. This feature also makes sense given the progress in constructing loop results from unitarity cuts, which use the physical on-shell amplitudes.

We have not used this running coupling in a calculation involving higher orders in the loop expansion, although we expect that this coupling remains an acceptable one. As mentioned in the introduction, in this case the renormalization group does not dictate the utility of the running coupling. However it appears likely that this coupling will appear in higher order processes.  Because of the simplicity of this theory, the $\phi^4$ interaction is the only one involved in higher order interactions. Within dimensionally regularization of loops for processes at an energy $E$, the only relevant momentum scale is again is the energy $E$ because the particles are massless. Dispersive techniques will use the on-shell amplitude, and hence will involve the coupling that we defined initially. Wick rotation of Feynman amplitudes would transform amplitudes to Euclidean momenta, where we found a similar result.

We conclude that this definition of a running coupling is a useful one in $\phi^4$ theory at one loop order and may also be useful at higher order in perturbative calculations.

%%%%%%%%%%%%%%%%%%%%%%%%%%%%%%%%%%%%%%%%%%%%%%%%%%%%%%%%%%%%%%%%%%
\section{Gravitational Corrections to Yukawa Interactions}
%%%%%%%%%%%%%%%%%%%%%%%%%%%%%%%%%%%%%%%%%%%%%%%%%%%%%%%%%%%%%%%%%%

In this section we follow the same lines above to calculate the $\beta$ function of Yukawa interactions. Here the basic vertex cannot be defined with all legs on-shell. Much as in gauge theory, on-shell renormalization requires a scattering amplitude with two vertices (the equivalent of Coulomb scattering in QED). However the off-shell function vertex can be defined at an unphyiscal kinematic point. We again explore both on-shell and off-shell renormalization.

If we define the original Yukawa coupling constant by the Lagrangian
\begin{equation}
{\cal L}_Y = \Gamma \phi {\bar \psi}\psi\,,
\end{equation}
then we are looking for a running coupling of the form
\begin{equation}
\Gamma (M)= \Gamma + a \Gamma \kappa^2 M^2
\end{equation}
for some constant $a$, when implemented with a renormalization scheme defined at the scale $M$. It will be possible to make such a definition.

However, we will also find that {\em any} such definition does not correctly capture the loop effects of the quantum corrections in all relevant processes. Let us highlight the main issue here before providing the explicit demonstration. Consider two physical processes involving the Yukawa couplings such as $f{\bar f}\to f {\bar f}$ and $f{ f}\to f { f}$ which proceed through the exchange of the scalar field. These of course are related by crossing, with the momentum-squared, $q^2$, of scalar changing from positive for the time-like process to negative for the space-like process. When calculated explicitly, the loop effects from the vertex function in these scattering processes will depend on the variable $q^2$. In this case the loop corrected matrix element will have the form
\begin{eqnarray}
{\cal M} &\sim& \left[\Gamma (M)\frac{1}{q^2}\Gamma (M) + {\rm loops}\right] \nonumber \\
&\sim& \left[ (\Gamma^2 + 2a \Gamma^2 \kappa^2 M^2 + 2 a' \Gamma^2 \kappa^2 q^2)\frac{1}{q^2}\right]\,,
\end{eqnarray}
where $a'$ is a number that emerges from the loop calculation. The problem is that even if the definition of the running coupling is chosen in such a way as to capture the main quantum corrections for one process, say $f{\bar f}\to f {\bar f}$, it will have the opposite effect in the crossed process. No definition of a running coupling can summarize the quantum corrections in both processes because the quantum effects go in different directions in the two cases. If they make the matrix element smaller in one channel, which naively looks like asymptotic freedom, they make the amplitude larger in the other process, which does not look like asymptotic freedom for the coupling.

On the other hand, operator mixing with a higher dimension operator does correctly  describe the quantum effects in both channels. Because the factors of $q^2$ cancel in the loop effects
\begin{equation}
(2 a' \Gamma^2 \kappa^2 q^2)\frac{1}{q^2} = 2 a' \Gamma^2 \kappa^2
\end{equation}
these effects are described by a contact operator
\begin{equation}
 a' \Gamma^2 \kappa^2{\bar \psi}\psi{\bar \psi}\psi\,.
\end{equation}
This works for both processes as the answer is independent of the sign of $q^2$.

\subsection{Operator mixing}

Gravitational corrections to the vertex (as shown in Fig. \ref{YukawaVertex}) will generally generate the higher order operators. A convenient basis for our calculation can be chosen to be
\begin{eqnarray}
\nonumber
{\cal O}_{1}&=&\phi\partial_{\mu}\bar\psi\sigma^{\mu\nu}\partial_{\nu}\psi\,,\\
\nonumber
{\cal O}_{2}&=&\phi\left(\bar \psi \partial^2\psi+\partial^2{\bar \psi} \psi \right)\,,\\
{\cal O}_{3}&=&\phi\partial_{\mu}\bar\psi\partial^{\mu}\psi\,.
\end{eqnarray}
and include their respective coupling constants
\begin{equation}
{\cal L}^{\rm h.o.} = g_1{\cal O}_{1} + g_2{\cal O}_{2}+g_3{\cal O}_{3}\,.
\end{equation}
Despite appearances, all three of these can be shown to vanish by the equation of motion. For ${\cal O}_{1}$ this follows from relating the operator to
\begin{equation}
{\cal O}_4= \phi\partial_{\mu}\bar\psi\gamma^{\mu}\gamma^{\nu}\partial_{\nu}\psi\,,
\end{equation}
which clearly vanishes by the Dirac equation, and then using the identity
\begin{equation}
\gamma_\mu \gamma_\nu = g_{\mu\nu} - i \sigma_{\mu\nu}\,,
\end{equation}
which follows from expressing the LHS in terms of commutators and anticommutators. This turns ${\cal O}_4$ into a combination of
other operators
\begin{equation}
{\cal O}_4 = {\cal O}_3 -i {\cal O}_1\,.
\label{operatorrelation1}
\end{equation}
In addition ${\cal O}_3$ can be seen to vanish through equations of motion through integration by parts. Here we define
\begin{equation}
 {\cal O}_5= \left(\partial^2 \phi \right) \bar\psi \psi\,,
\end{equation}
we find that
\begin{equation}
{\cal O}_5 = 2{\cal O}_3 +{\cal O}_2\,.
\label{operatorrelation2}
\end{equation}

However, a little more care is needed in this discussion because all legs of this three-point vertex cannot be on-shell at the same time. It is easy to see that an operator such as ${\cal O}_3 $ {\em can} lead to a non-vanishing matrix element in a physical process, such as $f{\bar f}\to f {\bar f}$, where the scalar field is off-shell. However, this matrix element is a constant independent of momentum, and is equivalent to a local four-fermion operator.    What this means is that another set of operators needs to be introduced which are used for onshell processes instead of the operators ${\cal O}_i$ which vanish by the equations of motion. These are
\begin{eqnarray}
{\cal Q}_{1}&=&\bar \psi \psi\bar\psi\psi\, \nonumber \\
{\cal Q}_2&=&\bar \psi\sigma^{\mu\nu}\psi \bar\psi\sigma_{\mu\nu}\psi \nonumber \\
{\cal Q}_{3}&=& \phi\partial_\mu \phi \bar \psi\partial^\mu\psi\,\nonumber \\
&& .....
\label{contactoperators}
\end{eqnarray}
Direct calculation using ${\cal O}_{i}$ in tree level physical processes shows that their effect are equivalent to contact operators such as the ${\cal Q}_i$ listed above. Equivalently one can use field redefinitions to remove the ${\cal O}_{i}$ operators, and these can generate the contact operators.

If we define the couplings of operators of Eq. \ref{contactoperators} by $q_i$, we will see that the $q_i$ will be infinitely renormalized by loop processes, and hence they are needed in the description of the on-shell renormalization.

%%%%%%%%%%%%%%%%%%%%%%%%%%%%%%%%%%%%%%%%%%
\subsection{Off-shell renormalization}
%%%%%%%%%%%%%%%%%%%%%%%%%%%%%%%%%%%%%%%%%%

\begin{figure}[ht]
\leftline{
\includegraphics[width=.45\textwidth]{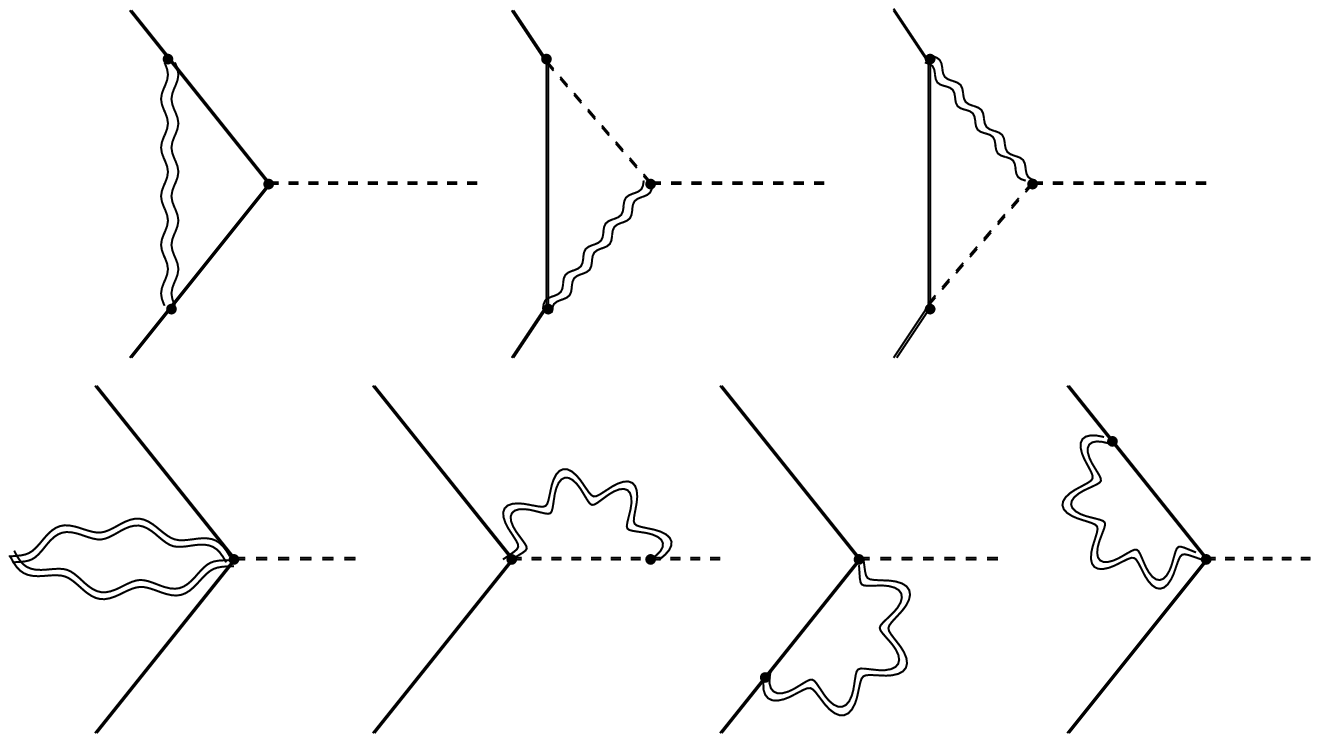}
}
\caption{Gravitational corrections to the vertex in Yukawa theory.}
\label{YukawaVertex}
\end{figure}

We first consider the vertex function directly and the quantum corrections to it as shown in Fig \ref{YukawaVertex}. Because we need to treat this vertex off-shell, we will need to include the unphysical operators that vanish by the equations of motion ${\cal O}_i$. We will see that it is possible to extract the effects of each of the three operators, using a set of renormalization conditions. Our treatment is relatively brief as the methods are similar to the off-shell renormalization of $\phi^4$ theory.

%%%%%%%%%%%%%%%%%%%%%%%%%
%%%%%%%%%%%%%%%%%%%%%%%%%%%%%%

In this method we use the Euclidean momenta $p_1^2=p_2^2=-\xi_1M^2$, and $p_1\cdot p_2=-\xi_2M^2$, where $p_1$ and $p_2$ are the fermions momenta, and $\xi_1$ and $\xi_2$ are arbitrary positive parameters. Unlike the case of $\lambda\phi^4$ theory, where we average over the different scattering channels, the loop results in the case of Yukawa theory will generally depend on the kinematics of the problem. In the following we parameterize this dependence using the parameter $\eta$. For time-like energy variable $q^2$ we have $\eta=1$, while for space-like $q^2$ we have $\eta=-1$, where $q=p_1+\eta p_2$.

\begin{figure}[ht]
\leftline{
\includegraphics[width=.45\textwidth]{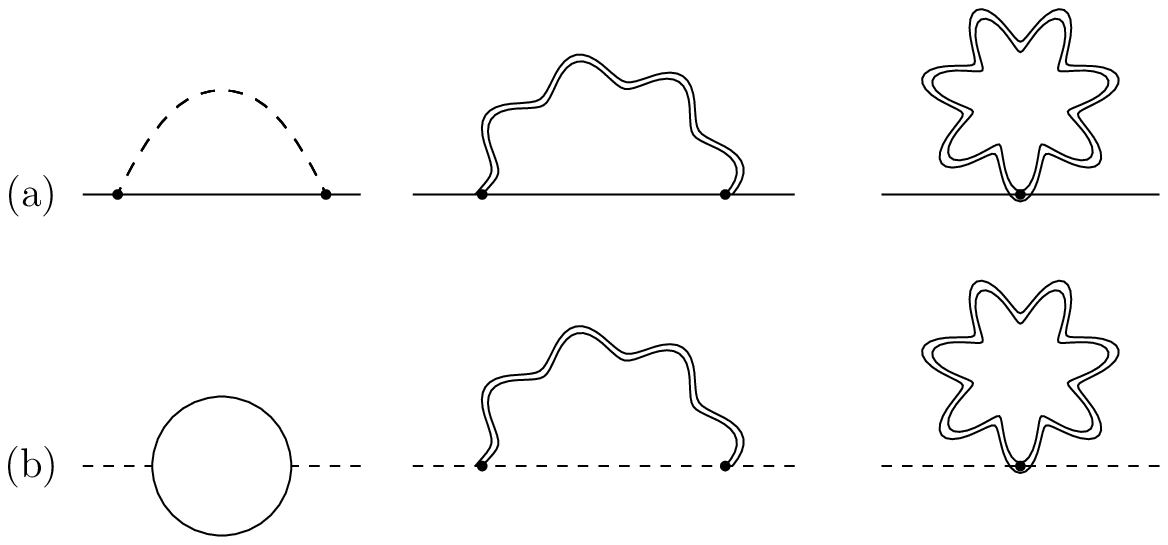}
}
\caption{ The diagrams contributing to the self energies of (a) fermions, and (b) bosons.}
\label{Yukawaself}
\end{figure}

In addition to the vertex diagrams shown in Fig. \ref{YukawaVertex}, we need to calculate the self-energy diagrams of Fig. \ref{Yukawaself}. The self energies and vertex corrections acquire logarithmic dependence of the form $\log\left(\xi_1M^2/\mu^2\right)$, and $\log\left((\eta\xi_2+\xi_1)M^2/\mu^2\right)$. We immediately see that only $\xi_1=0$, or $\xi_1=-\eta\xi_2$ are problematic, and hence we avoid these values in the following analysis.
We find that the total scattering amplitude, apart from finite pieces that do not affect the $\beta$ functions, is given by

%%%%%%%%%%%%%%%%%%%%%%%%%
%%%%%%%%%%%%%%%%%%%%%%%%%%%%%%

%
\begin{eqnarray}
\nonumber
{\cal A}&=&-ig+iM^2\left(2\xi_1 g_2-\xi_2g_3  \right)+g_1p_{1\mu}p_{2\nu}\sigma^{\mu\nu}\\
\nonumber
&&-\frac{g\kappa^2}{8\left(4\pi\right)^2}\left[-\frac{\eta\xi_1}{\xi_1-\eta\xi_2}{\cal S}_1+\frac{\xi_2}{\xi_1-\eta \xi_2}{\cal S}_2 \right]p_{1\mu}p_{2\mu}\sigma^{\mu\nu}\\
\nonumber
 &&+\frac{5ig^3}{2\left(4\pi\right)^2}{\cal S}_1-\frac{ig\kappa^2}{4\left(4\pi\right)^2}\left(\xi_1+\eta \xi_2\right)M^2{\cal S}_2\\
 \label{total scattering Yukawa}
&& +\frac{ig\kappa^2}{4\left(4\pi\right)^2}\left(\xi_1+2\eta \xi_2\right)M^2{\cal S}_1\,,
\end{eqnarray}
where
\begin{eqnarray}
\nonumber
{\cal S}_1&=&\left[\frac{2}{\epsilon}-\gamma+\log 4\pi-\log\left(\frac{\xi_1M^2}{\mu^2}\right)\right]\,,\\
\nonumber
{\cal S}_2&=&\left[\frac{2}{\epsilon}-\gamma+\log 4\pi-\log\left(\frac{2( \xi_1+\eta\xi_2)M^2}{\mu^2}\right)\right]\,.\\
\end{eqnarray}

In order to accomplish the renormalization we will define a set of renormalization conditions. We use the kinematic and Dirac structure to isolate the relevant terms. The overall amplitude is defined as
\begin{eqnarray}
\nonumber
{\cal A}&=&-ig(M)+iM^2\left(2\xi_1g_2(M)-\xi_2g_3(M)\right)\\
\label{yukawa definition of A}
&&+g_1(M)p_{1\mu}p_{2\nu}\sigma^{\mu\nu}\,.
\end{eqnarray}
By inspection, we find, apart from a trivial additive constant,
\begin{equation}
g_1(M)=g_1-\frac{\eta g\kappa^2}{8\left(4\pi \right)^2}\log\left(\frac{M^2}{\mu^2}\right) \,,
\end{equation}
where we have absorbed the pole $\sim p_{1\mu}p_{2\nu}\sigma^{\mu\nu}/\epsilon$ in $g_1$. We can also define $g_2(M)$ and $g_3(M)$ through the definitions
\begin{eqnarray}
\nonumber
\frac{\partial {\cal A}}{\partial (\xi_1 M^2)}=2ig_2(M)\\
\label{yukawa conditions}
\frac{\partial {\cal A}}{\partial (\xi_2 M^2)}=-ig_3(M)\,.
\end{eqnarray}
Hence, solving (\ref{yukawa definition of A}) and (\ref{yukawa conditions}) we find, apart from trivial additive constants,
\begin{eqnarray}
\nonumber
g(M)&=&g+\frac{5g^3}{2\left(4\pi \right)^2}\log\left(\frac{\xi_1M^2}{\mu^2}\right)-\frac{\eta\xi_2 g\kappa^2}{4\left(4\pi \right)^2}M^2\,,\\
\nonumber
g_2(M)&=&g_2-\frac{5}{4}\frac{g^3}{\left(4\pi \right)^2\xi_1M^2}\,,\\
\nonumber
g_3(M)&=&g_3+\frac{g\kappa^2\eta}{\left(4\pi\right)^2}\left[-\frac{1}{4}\log\left(\frac{2(\xi_1+\eta\xi_2)M^2}{\mu^2}\right)\right.\\
&&\left.+\frac{1}{2}\log\left(\frac{\xi_1M^2}{\mu^2}\right) \right]\,,
\end{eqnarray}
where the pole $\sim 1/\epsilon$ is absorbed in $g$, the pole $\sim M^2/\epsilon$ is absorbed in $g_3$, while $g_2$ does not get any pole contribution.
\footnote{This can be seen by taking the derivative of (\ref{total scattering Yukawa}) with respect to $\xi_1 M^2$.}
 Finally, the $\beta$ functions read
\begin{eqnarray}
\nonumber
\beta(g)&=&\frac{5g^3}{\left(4\pi\right)^2}-\frac{\eta\xi_2g\kappa^2}{2\left(4\pi\right)^2}M^2\,,\\
\nonumber
\beta(g_1)&=&-\frac{\eta g\kappa^2}{4\left(4\pi\right)^2}\,,\\
\nonumber
\beta(g_2)&=&\frac{5g^3}{\left(4\pi\right)^2\xi_1M^2}\,,\\
\beta(g_3)&=&\frac{\eta g\kappa^2}{2\left(4\pi\right)^2}\,.
\end{eqnarray}
Hence, we see the gravitational correction of $\beta(g)$ depends on the kinematics through the parameter $\eta$. More on this is discussed below.

%%%%%%%%%%%%%%%%%%%%%%%%%
\begin{figure}[ht]
\leftline{
\includegraphics[width=.45\textwidth]{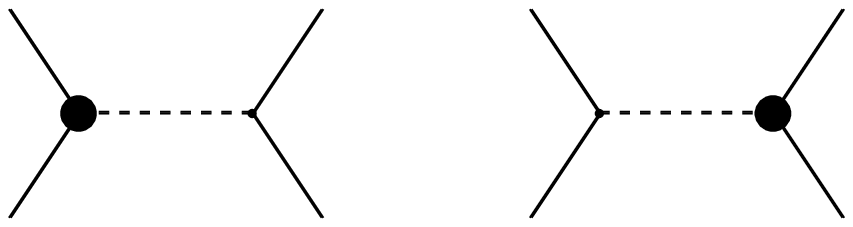}
}
\caption{Tree diagram for the on-shell scattering processes involving fermion.The filled circle denotes the set of vertex renormalization diagrams. }
\label{Treediagrams}
\end{figure}
%%%%%%%%%%%%%%%%%%%%%%%%%%%%%%

%%%%%%%%%%%%%%%%%%%%%%%%%%%%%%%%%%%%%%%%%%%%%%%%%%%%%%%%%%%%%%%%%%%%
\subsection{On-shell renormalization}
%%%%%%%%%%%%%%%%%%%%%%%%%%%%%%%%%%%%%%%%%%%%%%%%%%%%%%%%%%%%%%%%%%%%%

Finally we turn to the renormalization of the Yukawa coupling through an on-shell process. Similarly to the on-shell renormalization of the gauge couplings in QED, we consider the one-loop corrections to a scattering process such as $f+f \to f+f$ or $f+{\bar f} \to f+{\bar f}$. For clarity in separating the crossed channels, we will in this section refer to two flavors of fermions $f_a$ and $f_b$, so that we will compare $f_a+f_b \to f_a+f_b$ or $f_a+{\bar f}_a \to f_b+{\bar f}_b$ where the former has only a $t$-channel exchange and the latter only $s$-channel. Because these processes are on-shell, we can drop the operators ${\cal O}_i$ associated with the vertex itself, but must include the four fermion operators ${\cal Q}_1$ and ${\cal Q}_2$ associated with the four-fermion process.

\begin{figure}[ht]
\leftline{
\includegraphics[width=.45\textwidth]{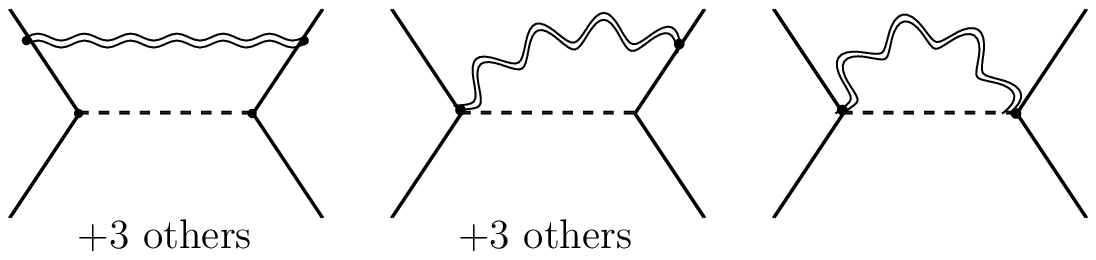}
}
\caption{Box diagrams.}
\label{Boxdiagrams}
\end{figure}

The key diagrams occur via the exchange of a scalar boson and so include the vertex correction on either side of the diagram, as in Fig. \ref{Treediagrams}. This set is the analogous to the set of diagrams considered for the running coupling in renormalizable theories. The process also includes a set of other diagrams, shown in Fig. \ref{Boxdiagrams}. While we have calculated the divergences in these diagrams and verified that they can be absorbed in the coefficients of the four fermion operators ${\cal Q}_i$, we do not include them in the definition of the running coupling.

%%%%%%%%%%%%%%%%%%%%%%%%%
%%%%%%%%%%%%%%%%%%%%%%%%%%%%%%

For this calculation we can use our previous results with the on-shell condition  $\xi_1=0$. We find that the gravitational corrections to the self energies vanish as it goes as $\sim \kappa^2p_i^2=0$. For the on-shell process $\bar ff\rightarrow \bar f f$ we find that the matrix element including gravitational corrections is given by
\begin{equation}
-i{\cal M}={\cal A}_{\scriptsize\mbox{on shell} -a}\frac{i}{\left(p_1+\eta p_2\right)^2}{\cal A}_{\scriptsize\mbox{on shell} -b}\,,
\end{equation}
where
\begin{eqnarray}
\nonumber
{\cal A}_{\scriptsize\mbox{on shell}-a}
&=&-ig_a-\frac{i\eta g_a\kappa^2}{\left(4\pi\right)^2}\left[ -\frac{3}{8}S_2\right.\\
\label{on shell Yukawa amplitude}
&&\left.+\frac{1}{2} S_1-\frac{3}{16}\right]p_1\cdot p_2\,,
\end{eqnarray}
and
\begin{eqnarray}
\nonumber
{\cal S}_1&=&\left[\frac{2}{\epsilon}-\gamma+\log 4\pi-\log\left(\frac{m_{\psi}^2}{\mu^2}\right)\right]\,,\\
\nonumber
{\cal S}_2\left( p_1\cdot p_2\right)&=&\left[\frac{2}{\epsilon}-\gamma+\log 4\pi-\log\left(\frac{-2\eta p_1\cdot p_2}{\mu^2}\right)\right]\,.\\
\end{eqnarray}
In the above analysis we have restored the fermion mass in order to avoid an IR singularity in the logarithm. Notice that the result (\ref{on shell Yukawa amplitude}) can be obtained from (\ref{total scattering Yukawa}) by setting $\xi_1=0$ everywhere except inside the logarithm where it is replaced by the fermion mass, and replacing $M^2\rightarrow -M^2$, i.e. working with physical rather than Euclidean momenta.  Hence,
\begin{eqnarray}
\nonumber
&&-i{\cal M}\left( p_1\cdot p_2=\xi_2 M^2 \right)=-\frac{ig_ag_b}{2\eta\xi_2M^2}\\
\nonumber
\label{final form for on shell matrix}
&&~~~~~~~~~~+\left\{\frac{ig_ag_b\kappa^2}{2\left(4\pi\right)^2}\left[\frac{3}{8}S_2(\xi_2M^2)-\frac{1}{2}S_1+\frac{3}{16} \right]\right.\\
\nonumber
&&~~~~~~~~~~\left.+\left(a \leftrightarrow b\right)\right\}\,,\\
\label{final form for on shell matrix}
\end{eqnarray}
and we have used $p_1\cdot p_2=\xi_2 M^2$.

In order to use this result for the renormalization of $g$, we define the $t$-channel process $f_a+f_b\to f_a+f_b$ with our kinematic features defined by $\eta=-1, ~\xi_2=1/2$, $q^2 =-2p_1\cdot p_2= -M^2$ defining the matrix element
\begin{eqnarray}
\nonumber
-i{\cal M}(f_a+f_b\to f_a+f_b) = (-ig_a^{r} ) \frac{i}{q^2} (-ig_b)\\
+\left(a \leftrightarrow b\right)   -i q_1\,.
\end{eqnarray}
Here we have included only the coefficient of the four-fermion operator ${\cal Q}_1=\bar \psi_a \psi_a \bar \psi_b \psi_b$, as described in Eq. \ref{contactoperators}. The other operators are not needed for this analysis.

 We can absorb the pole in (\ref{final form for on shell matrix}) $\sim 1/\epsilon$, which is due to gravitational corrections, into $q_1$ by defining
\begin{equation}
q_{1}^r=q_1-\frac{g_ag_b\kappa^2}{\left(4\pi\right)^2}\left[\frac{3}{8}S_2 \left(\frac{M^2}{2}\right)-\frac{1}{2}S_1 \right]\,.
\end{equation}
The residual loop effects are contained in the renormalized value of $g$ and we find
\begin{equation}
g_i^r =g_i +\frac{3g_i\kappa^2M^2}{32(4\pi)^2}
\end{equation}
for $i=a,b$.

At this point, one can calculate the gravitational contributions to the $\beta$ function of $g$ and $q_1$ as
\begin{equation}
\beta(g_i)\equiv M\frac{\partial g_i^r}{\partial M}=\frac{3g_i\kappa^2M^2}{16(4\pi)^2}\,.
\label{beta function Yukawa}
\end{equation}
and
\begin{equation}
\beta(q_1)\equiv M\frac{\partial q_{1}^r}{\partial M}=\frac{3g_ag_b\kappa^2}{4(4\pi)^2}\,.
\end{equation}

\subsection{The problem with the running Yukawa coupling}

We have been able to define a coupling constant at the scale $M$ by direct calculation of the process
$f_a+f_b\to f_a+f_b$ with the scalar exchange in the $t$ channel. It works as expected when dealing with this process near the renormalization point.
Explicitly we see that
\begin{eqnarray}
\nonumber
&&-i{\cal M}(f_a+f_b \to f_a+f_b)(q^2\sim M^2) = \\
\nonumber
&&~~~~\left(-ig_a^{r}(M^2) \right) \frac{i}{q^2} \left(-ig_b^{r}(M^2)\right)  -i q_1^r(M^2)\\
\nonumber
 &&~~-i\frac{3g_ag_b\kappa^2 M^2}{16(4\pi)^2}\left(\frac{1}{|q^2|}-\frac{1}{M^2}\right) -i\frac{3g_ag_b\kappa^2}{8(4\pi)^2}\log \left(\frac{|q^2|}{M^2}\right)\,.\\ \end{eqnarray}
Here we see that including the some of the loop effects in the coupling constant at the scale $M^2$ does a reasonable job of capturing the quantum effects in the neighborhood of this point. By construction, the residual quantum effects vanish if we choose $q^2=-M^2$

However, now we consider the process $f_a+{\bar f}_a\to f_b+{\bar f}_b$ with the scalar exchange in the $s$ channel, with $s=(p_1+p_2)^2>0$, we see that that this definition of the running coupling has the wrong sign to correctly capture the effect of the quantum loops. Instead of incorporation the majority of quantum effects at this scale gets the sign wrong so that the quantum effects are doubled.
Explicitly we find
\begin{eqnarray}
\nonumber
&&-i{\cal M}(f_a+{\bar f}_a\to f_b+{\bar f}_b)(s\sim M^2) = \\
\nonumber
&&~~~~~~~\left(-ig_a^{r}(M^2) \right) \frac{i}{s} \left(-ig_b^{r}(M^2)\right)  -i q_1^r(M^2) \\
\nonumber
&&~~+i\frac{3g_ag_b\kappa^2 M^2}{16(4\pi)^2}\left(\frac{1}{s}+\frac{1}{M^2}\right) -i\frac{3g_ag_b\kappa^2}{8(4\pi)^2}\log \left(\frac{-s}{M^2}\right)\,.\\
\end{eqnarray}
The process of absorbing these $\kappa^2 M^2 $ effects in the coupling constant has not described the kinematic dependence of the loops in this crossed reaction because the sign of $q^2$ has changed. The residual power-law effects are doubled at $s=M^2$, while the log effects associated with the running of $q_1$ behave as expected.

Alternatively we could have used the process $f_a+{\bar f}_a\to f_b+{\bar f}_b$ as our choice of renormalization procedure. In this case, the quantum correction and the running of the coupling $g$ in the beta function Eq. \ref{beta function Yukawa} would have the opposite sign. This defintion would work fine for the $s$-channel process, but would then fail for the $t$-channel reaction.

It is clear that we can have a sensible on-shell renormalization procedure which yields either sign for the running of the coupling, depending on whether we use the space-like or time-like reaction. Indeed there is a multiplicity of schemes that can generate a wide range of answers. It is also clear from this that no single scheme (either on-shell or off-shell) will be able to correctly categorize the leading quantum effects in all processes, because the crossed reaction will have a result of the opposite sign. This is to be contrasted with the definition of running coupling in renormalizable theories, which is universally successful for all reactions independent of the renormalization scheme.

The difference comes from the phenomenon of operator mixing. In the case of renormalizable theories, the quantum effects are renormalizing the original operator. However in effective field theories the quantum effects are associated with an operator of higher dimension. We can by fiat take some of the higher order effect and build it into the original coupling - there are many ways to do this. However, we see that such a construction fails to account for the quantum effects in related processes in the correct way.

There are further process dependence that is evident in our results. If we were to consider a different on-shell reaction, say $\phi +\psi \to \phi +\psi$, this would involve the basic $\phi{\bar \psi}\psi$ vertex with a different particle off-shell, in this case the fermion. Our result for the off-shell renormalization shows that this involves a different numerical factor, and equivalently the overall process involves different operators in the ${\cal Q}_i$ basis of Eq. \ref{contactoperators}. Again we see that because the loops renormalize higher order operators that do not lead to a universal running coupling for all processes.

%%%%%%%%%%%%%%%%%%%%%%%%%%%%%%%%%%%%%%%%
\section{Brief comments on the literature}
%%%%%%%%%%%%%%%%%%%%%%%%%%%%%%%%%%%%%%%%%

The study of gravitational corrections to the running of gauge couplings was started by Robinson and Wilczek \cite{Robinson:2005fj} who calculated the one-loop contribution of graviton exchange to the $\beta$ function of the Yang-Mills theory using the background field method. Subsequently a series of authors  \cite{Pietrykowski:2006xy, Toms:2007sk, Ebert:2007gf} used a variety of different methods to argue that the gravitational correction to the running of gauge couplings actually vanishes. Yet more recently, there have been further studies \cite{Tang:2008ah, Daum:2009dn, Toms:2010vy, He:2010mt} that again claim non-zero effects for the running of gauge couplings. Similar discrepancies are found in work that studies the non-gauge interactions of scalars and fermions
\cite{Griguolo:1995db,Zanusso:2009bs, Rodigast:2009zj,Mackay:2009cf, Shaposhnikov:2009pv}
again using a variety of methods.

The variety of answers found using different methods is itself a indication of the non-universality of gravitational corrections to running couplings. Because of the power counting for gravitational loop corrections, the true effect of gravitational loops is to renormalize a higher order operator and not the original vertex. Of course, in some schemes with a dimensionful cutoff, there is an additional renormalization of the original operator that is not seen when using dimensional regularization. However, this is a scheme dependent artifact and we comment on this below.

In addition, most schemes in the literature do not consider the effects of high order operators nor of process dependence. By mimicking the effect of higher order operators through a redefinition of the original coupling, one can create the appearance of a running coupling. But unless this definition is universally valid, it is not a true running coupling. Our work shows the limitations of such definitions. Moreover, as we have shown, the operator basis plays a different role in off-shell methods versus on-shell methods.

%%%%%%%%%%%%%%%%%%%%%%
\subsection{Comments on cutoff regularization}
%%%%%%%%%%%%%%%%%%%%%

We have performed our calculations using dimensional regularization. Real physics does not depend on the renormalization scheme, so that we could obtain equivalent results in any consistent regularization scheme that respects general covariance. However, because the gravitational coupling is dimensionful, the cutoff can appear as a power in the renormalization procedure and this can create some confusion.

Let us review how one would recover the results of dimensional regularization in situations in which a cutoff is used. One imagines that the momentum range is divided into regions above and below an arbitrary cutoff $\Lambda$. The contribution of the region above the cutoff is of course unknown, and the effects are contained in the coupling constant $g^{\rm bare}(\Lambda)$ which is parameterized by the separation scale $\Lambda$. The theory below the cutoff is treated as a full field theory and the loops are cutoff by $\Lambda$. In the case of gravity this cutoff appears quadratically because of the dimensional gravitational coupling. The bare coupling and the loop correction are added together to yield the renormalized physical coupling
\begin{equation}
g^{\rm phys.} = g^{\rm bare}(\Lambda) + c \kappa^2 \Lambda^2
\end{equation}
which however is independent of the arbitrary cutoff $\Lambda$. It is the physical coupling which is equivalent to the coupling used in a dimensionally regularized scheme. In dimensional regularization, the loop momentum runs over all energies and there is no need for a separation scale, nor for a divergence in the renormalization of the original operator. As expected then, physical effects are the same with either regularization scheme.

Should one identify the dependence of the bare coupling  $g^{\rm bare}(\Lambda)$ on the cutoff as the running of the coupling constant? The answer is clearly negative. That dependence cancels completely in physical observables. It also is independent of the kinematic dependence of scattering amplitudes so that it does not capture the real quantum effects of physical processes. This situation is different in the case of renormalizable field theories. In those cases the dependence on the cutoff is logarithmic, and by dimensional grounds the kinematic variables also enter into the logarithm, $\ln (\Lambda^2/q^2)$. So even though the cutoff cancels in physical observables it nevertheless provides a guide to the kinematic dependence of the quantum corrections. The counterpart in dimensional regularization is the study of the $1/\epsilon$ and $\ln (\mu^2/q^2)$ terms which also mirror the kinematic dependence of loop effects. The fact that $1/\epsilon$ or  $\ln (\Lambda^2)$ goes into the renormalization of the coupling also leads to the universality of these quantum effects, independent of the process being considered.

We see that the quadratic cutoff dependence in gravitational corrections does not signal the appearance of a running coupling in physical processes but totally drops out as an unphysical artifact. This is why we focussed directly on the kinematic dependence of physical processes and the influence of that on a coupling through the choice of renormalization procedure. This type of behavior is independent of the regularization scheme.

%%%%%%%%%%%%%%%%%%%
\section{Discussion}
%%%%%%%%%%%%%%%%%%%%

Effective field theory techniques expand a theory about zero energy, and the renormalization and higher momentum dependence involves the operators that appear at higher order in the energy expansion. Here the content of the renormalization group is limited to the description of the coefficients of the leading logarithmic loop corrections, but the couplings of the theory do not run with the energy scale. In contrast, there are many attempts to define running coupling constants in such theories by renormalizing the theory at a higher energy scale $M$. We have explored this process using off-shell and on-shell renormalization techniques, and have studied the subsequent utility of a running coupling in perturbative scattering processes. This provides a well-defined setting for exploring the nature of these running couplings.

It is of course possible to define running couplings in effective theories such as gravity. Indeed, there are a quasi-infinite number of ways of defining such couplings, so the question is not whether it can be done but rather whether it is useful and also to what extent it is universal. The comparison standard is the definition of running couplings in renormalizable theories where, despite minor scheme dependence, the results are highly useful and universal.

Unfortunately, the corresponding definitions including gravity are in general seen not to be useful or universal when describing physical reactions. The corrections go like $\kappa^2 q^2$, and in physical processes $q^2$ can be either positive or negative. For theories such as the Yukawa couplings, it is not possible to define a running coupling that is appropriate for both space-like and time-like processes. A theory such as $\phi^4$ is seen to be an exception. Gauge theories and gravity coupled to matter would behave similar to the Yukawa theory.

Instead, we see that in the perturbative realm the physics which would be described by a running coupling is accounted for by operator mixing. Higher dimension operators with derivative coupling appropriately deal with both space-like and time-like processes.

In our studies we also demonstrated how operators which vanish by field redefinition or by the equations of motion, such that they would normally be discarded from the operator basis, are nevertheless required when one performs the renormalization at an off-shell point. This is not surprising because the general theorems which demonstrate the irrelevance of such operators only apply on-shell. For this reason the effects of these operators continues to disappear when applied to physical processes. However, it points to the need of an expanded operator basis for field theory methods that involve off-shell renormalization.

\section*{Acknowledgements}

M.A. would like to thank T.J. Blackburn for helpful discussions. The work of J.D. is supported in part by the U.S. NSF grant PHY-0855119. The work of  M.A. has been supported in part by NSERC Discovery Grant of Canada, and in part by the U.S. NSF grant PHY-0855119. The work of  M.E. has been supported by Mission Department, the Egyptian Ministry of Higher Education under the scheme Data Collection Program.

\end{document}